\documentclass[11pt,a4paper]{article}
\usepackage{eepic}
\usepackage{epsf}
\usepackage{amsmath}
\usepackage{graphicx}
\usepackage[normalem]{ulem}
\usepackage{hyperref}
\usepackage{lipsum}
\usepackage[superscript]{cite}
\newcommand \bea {\begin{eqnarray}}
\newcommand \eea {\end{eqnarray}}

\begin{document}
\baselineskip=12pt
\begin{center}
{\LARGE{Designing intramolecular singlet-fission materials using indeno[1,2-b]fluorene dimers: A
DMRG and TDDFT study.}}\\
\vspace*{1.0cm}
Sumit Naskar$^1$, Mousumi Das$^{*1}$ \\
$^1${\it Department of Chemical Sciences, Indian Institute of Science Education and Research Kolkata, 
Mohanpur - 741246, India, email: mousumi@iiserkol.ac.in}\\
\end{center}

\vspace*{2.0cm}
\begin{abstract}
Low-lying excited states for indeno[1,2-b]fluorene homo dimers with or without
benzene spacers are calculated using the Density Matrix Renormalization group
(DMRG) approach within Pariser-Parr-Pople (PPP) model Hamiltonian. DMRG 
calculations suggest that all the dimers studied here satisfy the essential
energy conditions for SF. SF is a multiexciton generation process. As it is 
spin allowed, the process is very fast. By generating multiple exciton at a
time SF underestimate SQ limit to enhance photo-conversion efficiency of single 
junction solar cells. Frontier orbital calculation through 
Density Functional Theory (DFT) depicts orbital localization of triplets on 
either side of the covalent spacers. Which supports the entangled 
triplet-triplet state $^1(TT)$. Here the process is intramolecular (iSF), 
which has many advantages over the intermolecular (xSF) process, as in 
intermolecular process the SF process is highly dependent on the crystal 
packing, defects, dislocations etc. The entangled $^1(TT)$ state for xSF is 
localized on both of the chromophores, thus the appropriate crystal packing is 
essential for xSF. However iSF does not depend on the crystal packing. Our
DMRG calculation and TDDFT calculation are in well agreement with experimental
results found in the literature. Thus indeno[1,2-b]fluorene homo dimers can be
applicable in iSF application.
\end{abstract}
\baselineskip=24pt

\newpage
\section{INTRODUCTION}
\begin{figure}
\begin{center}
\includegraphics[width=1.0\linewidth]{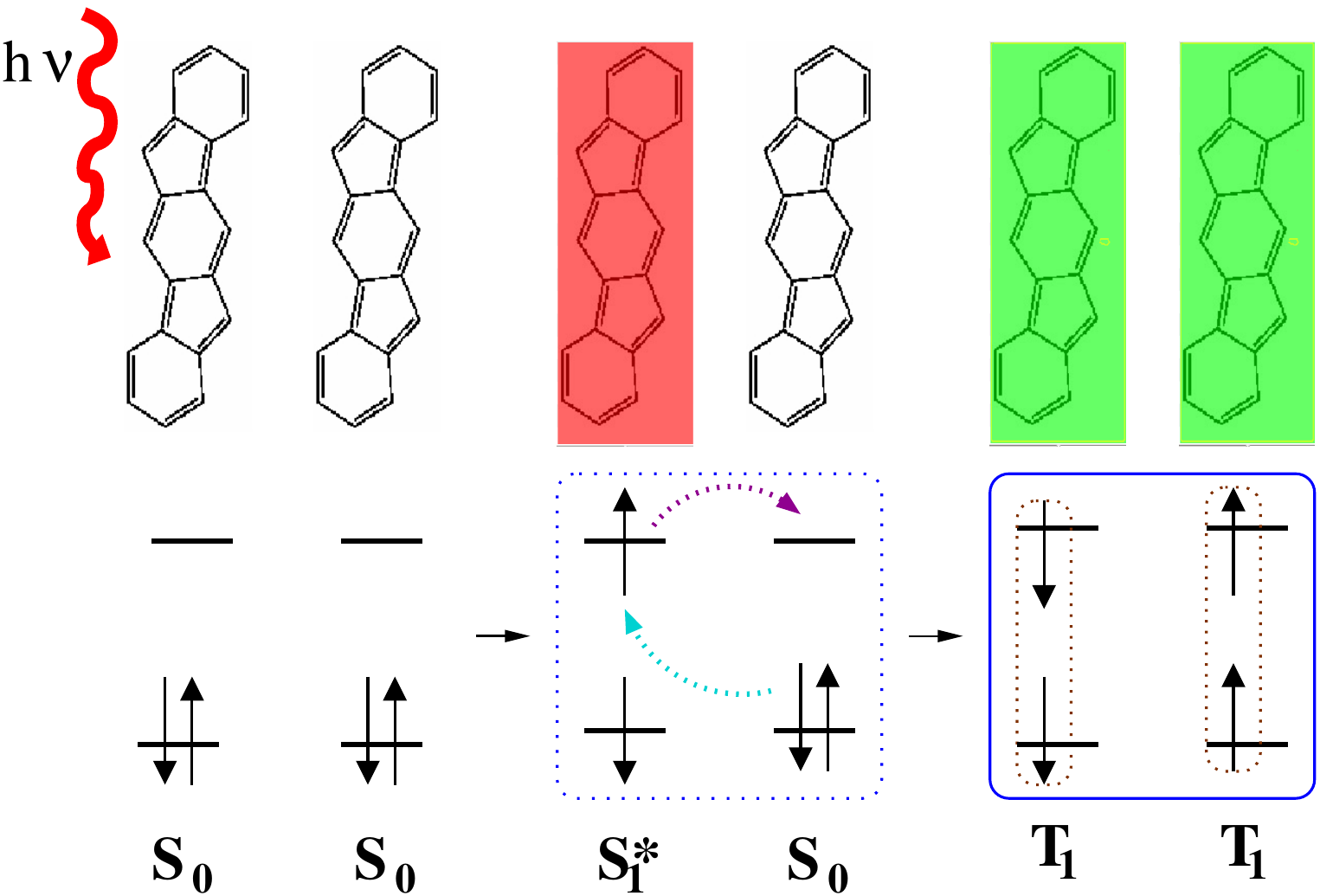}
\caption{\bf{Schematic diagram of xSF}}
\label{xSF-schematic}
\end{center}
\end{figure}

Indeno[1,2-b]fluorene (IF) is a conjugated, antiaromatic 4n$\pi$ polycyclic hydrocarbons system contains 20 Carbon atoms
 and it has singlet diradical character at ground state \cite{tobe2018quinodimethanes,lopez2017singlet,fix2012indenofluorenes,ito2016diradical}. 
This kind of polycyclic hydrocarbon systems are rigorously studied due to their robust molecular structure\cite{frederickson2016modulating}. 
Recently these kind of systems are in the center of interest due to their opto-electronic properties and device applications.\cite{chaurasia2015sensitizers,fukuda2017theoretical}
 One interesting optical property of IF is intermolecular singlet-fission (xSF) (schematically shown in Fig\ref{xSF-schematic}) in which one excited 
singlet exciton produces two or more excited triplet excitons through a charge transfer state $^1(TT)$ and satisfies the 
energy criteria given by Paci et al.\cite{paci2006SF} as:
\begin{eqnarray}
E(S_{0}-S_{1})\ge 2E(S_{0}-T_{1})
\end{eqnarray}
and
\begin{eqnarray}
E(S_{0}-T_{2})> 2E(S_{0}-T_{1})
\end{eqnarray}
The first equation is necessary energy condition for SF and the 2nd equation causes low and slow recombination of triplet
pairs. The newly generated entangled triplet pair $^1(TT)$ can be annihilated in three possible ways\cite{paci2006SF}:
\begin{enumerate}
\item The pair can be annihilated to form $S_{0}+S_0$ state and the process is very exoergic and slow.
\item During fission there can be dephasing between the triplet pair which can be very fast. Due to this dephasing there will
not be any correlation between the triplets, thus annihilation of them can cause a overall triplet or a quintet (Q) state.
Now annihilation to produce a $Q_{1}+S_0$ state is very much endoergic and yield of $T_{1}+S_0$ requires exoergic transfer
of the triplet excitons.
\item Only annihilation to yield $T_2$ and $S_0$ is almost isoergic and very fast.
\end{enumerate}
That is why the 2nd condition for SF occurred as if the $^1(TT)$ is almost isoergic to next higher excited triplet state
then the pair can be annihilated to form $T_2$, which is not desirable for the long lifetime of the triplet pairs.\\

\begin{figure}
\begin{center}
\includegraphics[width=0.8\linewidth]{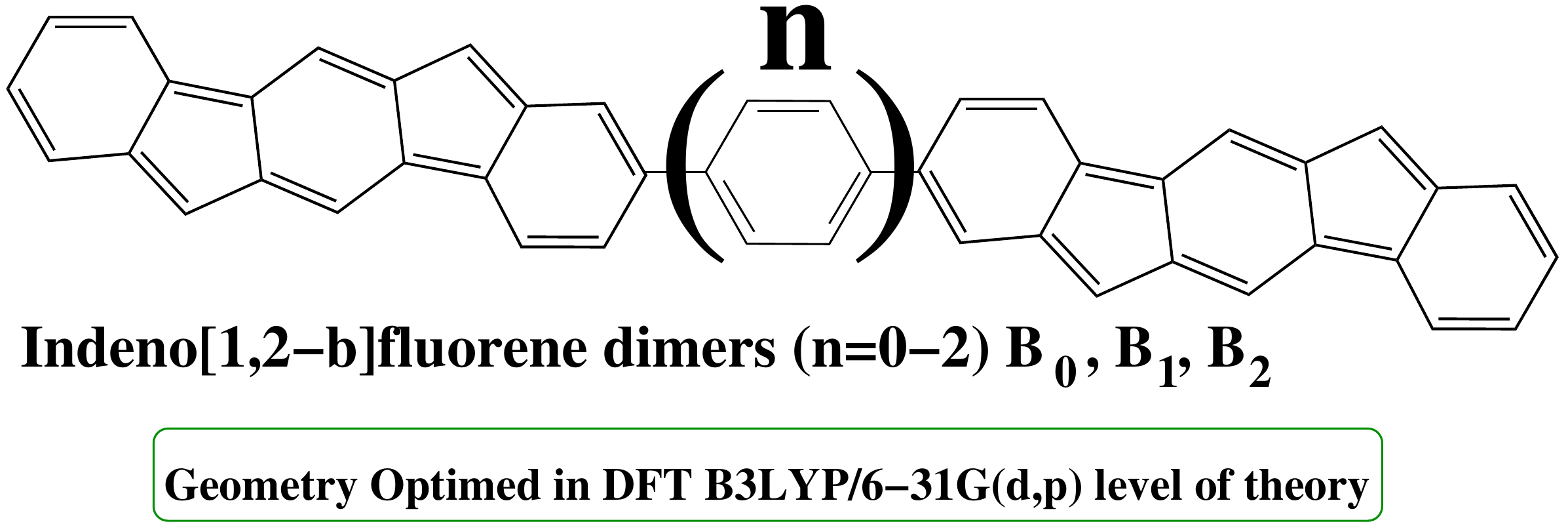}
\caption{\bf{Indeno[1,2-b]fluorene dimers B0, B1, B2}}
\label{molecules-schematic}
\end{center}
\end{figure}

\begin{figure}
\begin{center}
\includegraphics[width=1.0\linewidth]{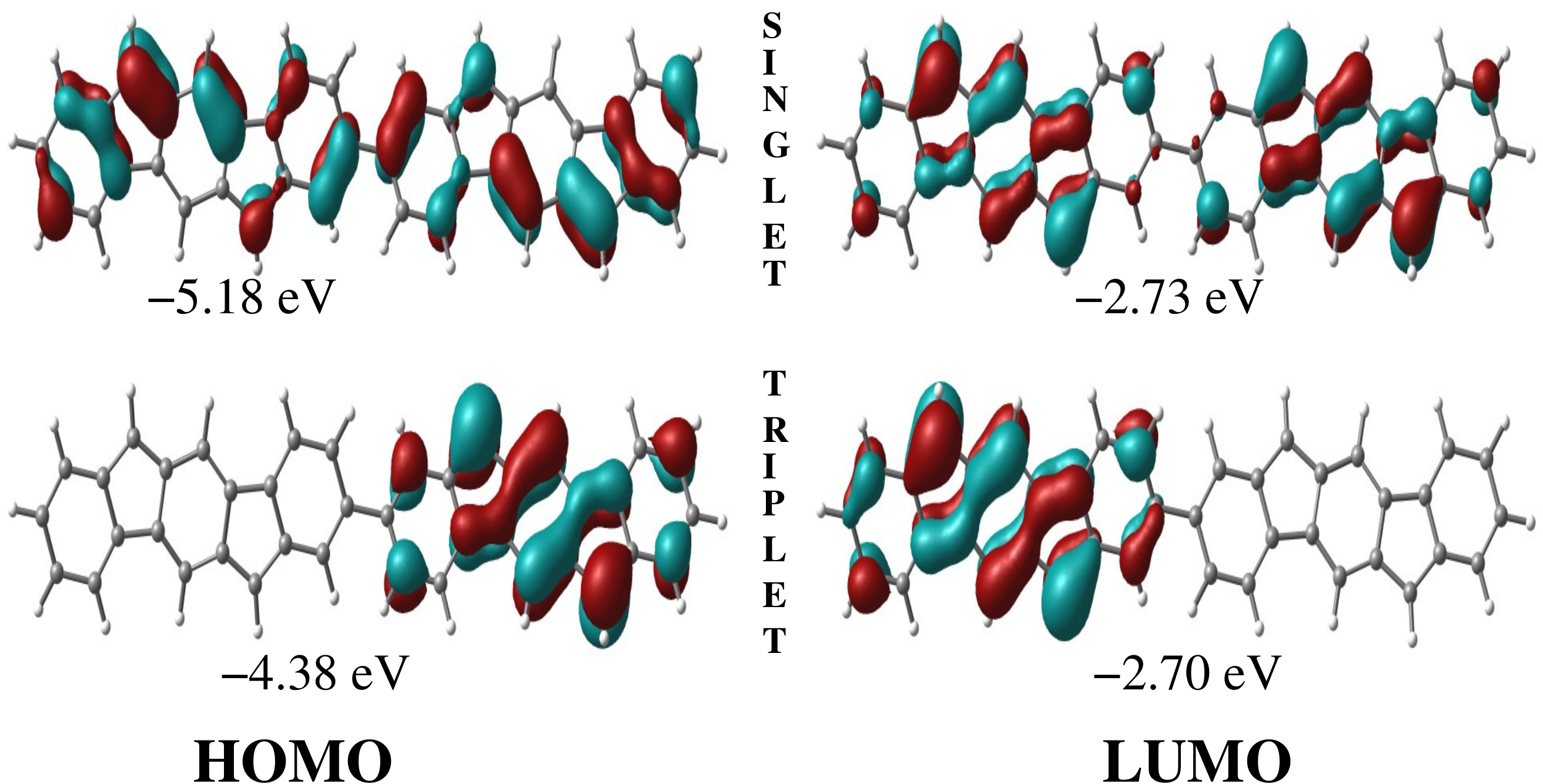} 
\caption{\bf {Frontier orbitals of B0 singlet ground state and triplet state}}
\label{indeno-B0}
\end{center}
\end{figure}

\begin{figure}
\begin{center}
\includegraphics[width=1.0\linewidth]{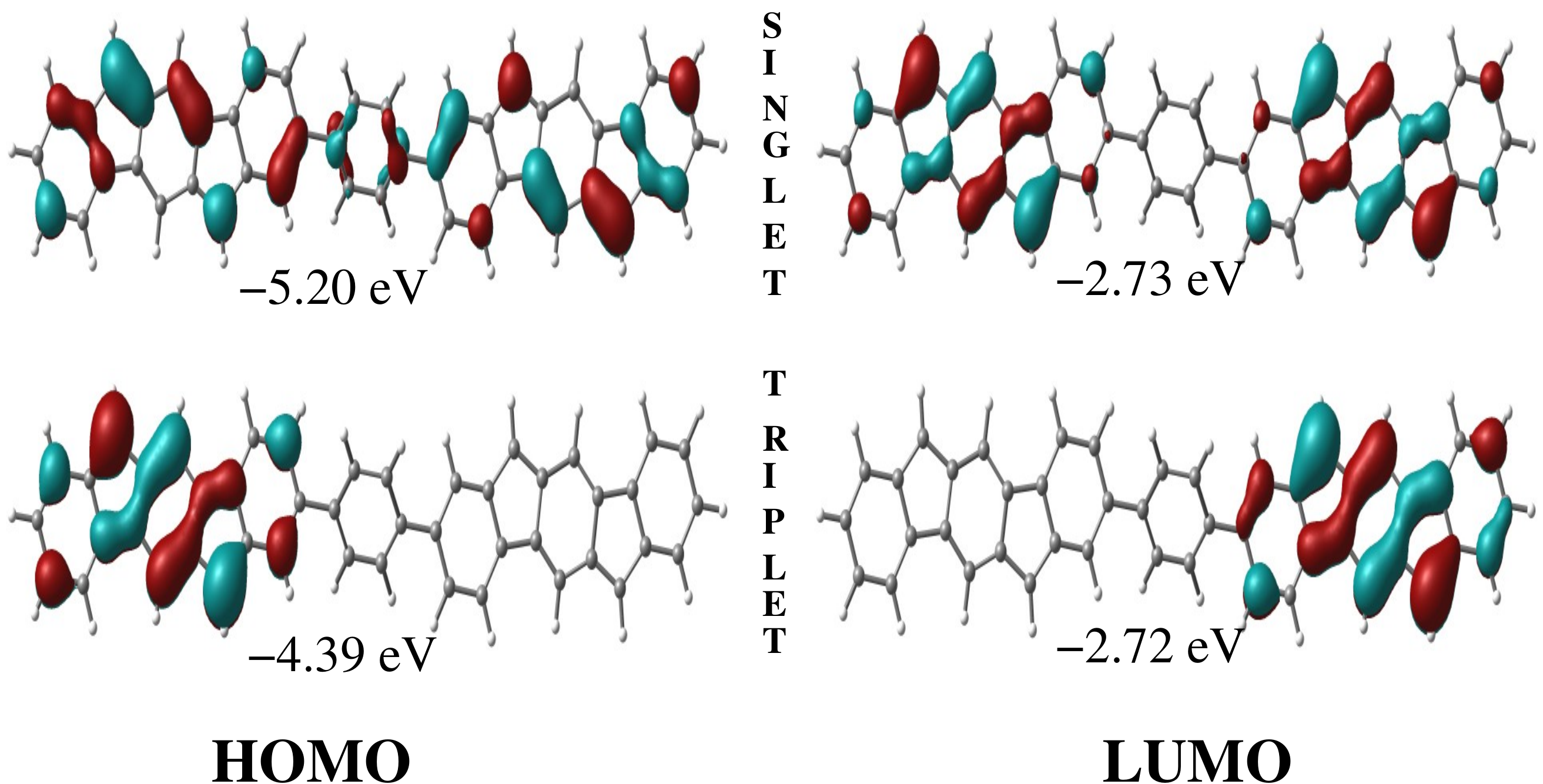} 
\caption{\bf {Frontier orbitals of B1 singlet ground state and triplet state}}
\label{indeno-B1}
\end{center}
\end{figure}

\begin{figure}
\begin{center}
\includegraphics[width=1.0\linewidth]{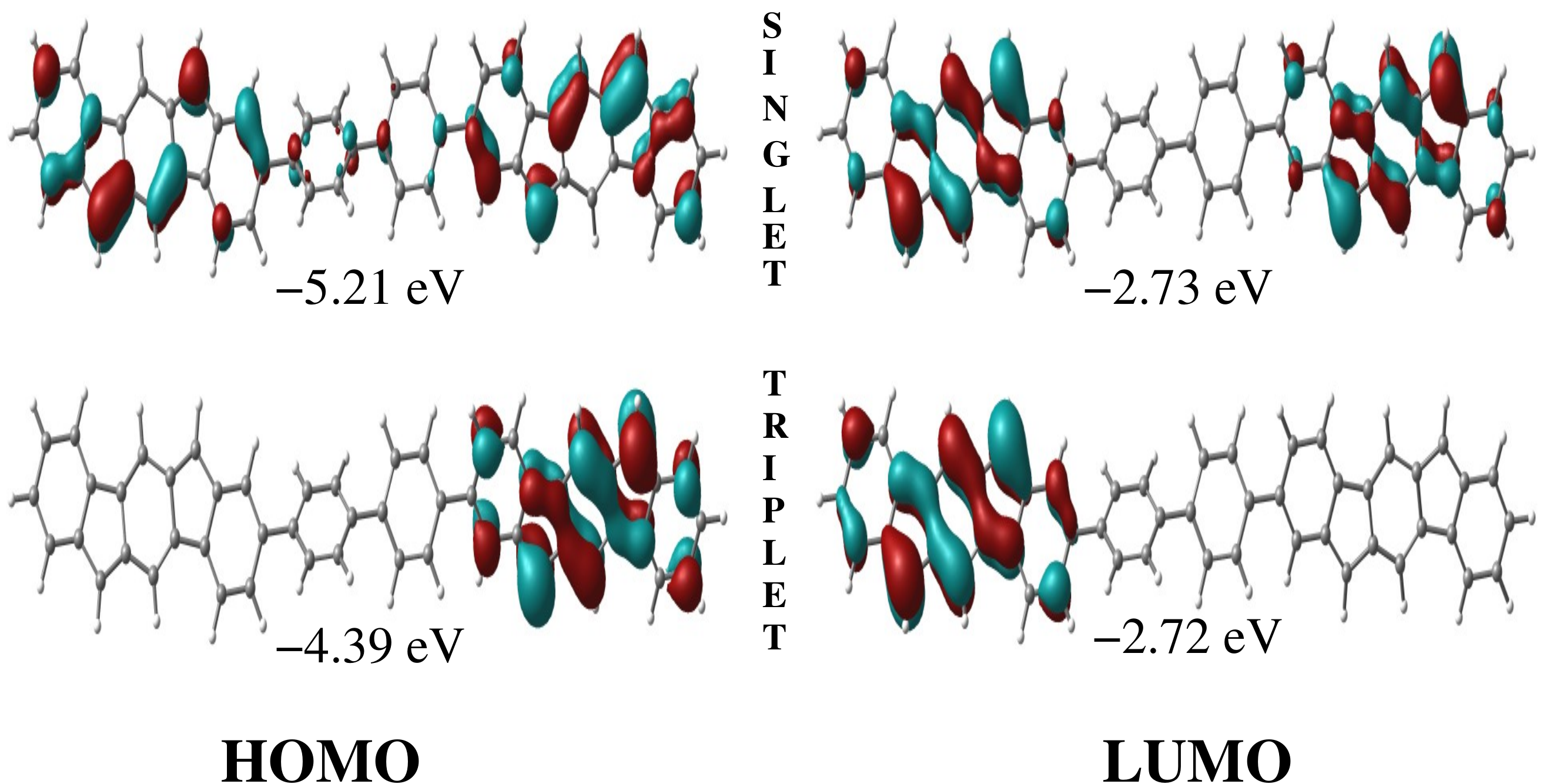} 
\caption{\bf {Frontier orbitals of B2 singlet ground state and triplet state}}
\label{indeno-B2}
\end{center}
\end{figure}

As the triplets are in the odd spin parity symmetry the SF process is spin allowed and thus quite fast.
This SF phenomenon is so strong that it can be used as the quantum efficiency enhancer of single junction solar cells by
modulating the theoretical limit of photo conversion efficiency known as Shockley-Queisser (SQ) limit. It can be used as
active catalyst as it can generate multiple electrons at a time. However the xSF pheomenon is depends on the tansfer of 
charge to the nearest neighbour. Thus the chromophore should be densely packed without any defect. But in practice there
would be defects and dislocations and structural heterogeneity. Thus molecules with xSF are found to be difficult in solid
phase to produce SF mechanism although one single molecule satisfies the energy criterion given in equation (1) and (2). 
The possible way out is intramolecular singlet-fission (iSF). In this process there are molecular dimers rather than the 
monomers and the dimers are made by the monomers with or without covalent spacers. The yield of these triplet pair is 
$\sim 30\%$ before 2015. In 2015 Zirzlmeier et al. reported the triplet yield of 156\% and Samuel N. Sanders et al. 
reported the yield of 170\% \cite{zirzlmeier-pnas,sanders-jacs}. The iSF mechanism found by different group occurred in 
dilute solution phase rather 
than solid crystalline phase or dense solution phase, which almost completely remove the structural heterogeneity and defects 
occurred during xSF process. iSF process is governed by the localized triplet states on the molecular fragment, which came 
from the diffusion of excited triplets of one fragment of the dimer. 
In Fig \ref{indeno-B0}, \ref{indeno-B1} and \ref{indeno-B2} it can be clearly shown that the
coupled $^1(TT)$ state are localized on the optimized triplet geometry of the diners.\\
Here in our study we choose equivalently linked indent[1,2-b]fluorine diners with (oligopoly)f-
ethylene spacers (n=0,1,2) 
(Fig\ref{molecules-schematic}) named as B0, B1, B2.
\begin{figure}
\begin{center}
\includegraphics[width=1.0\linewidth]{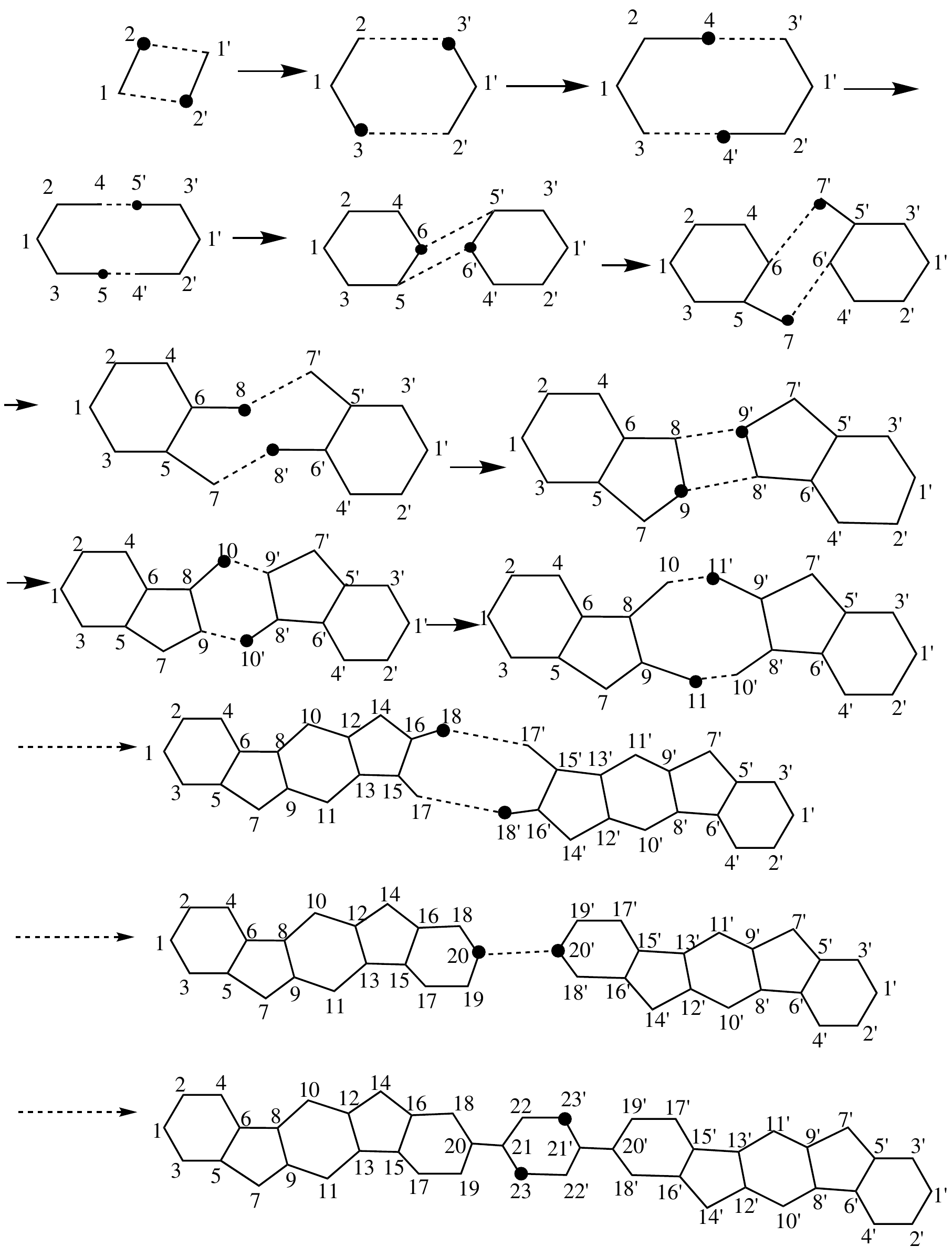}
\caption{\bf {Connectivity scheme of covallently linked indeno$[1,2-b]$fluorene dimer}}
\label{iSF_connect}
\end{center}
\end{figure}
The two-photon absorption levels ($2A_g$) are below the $S_{0}-S_{1}$ level observed for all the dimers studied. Thus these
systems shows exoargic transfer of excitons as it is observed theoretically in the monomer. Figure 3, 4 and 5 shows the
delocalised singlet state of B0, B1 and B2 all over the molecules as it was previously found for the 
triisopropysilylacety-lene (TIPS) substituted pentacene dimer reported by Samuel N. Sanders et al and it is differed
from the report from Zirzlmeier group\cite{zirzlmeier-pnas,sanders-jacs}.

\clearpage
\section{COMPUTATIONAL APPROACH}
Conjugated systems follows 4n and 4n+2 H$\ddot{u}$ckel rule, i.e they have 4n or 4n+2 
no. of $\pi$ electrons in the $p_z$ orbital while the $\sigma$ electrons 
form $sp^2$ hybridized backbone. Earlier these interacting $\pi$-electronic
 systems are described by Pariser-Parr-Pople model (PPP) Hamiltonian 
as the Hamiltonian (PPP) physically and numerically depicts the $\pi$ 
electronic configuration at the valence orbitals of these systems.\cite{pariser1953-I,pariser1953-II,pople1953}
 Which includes long range hopping of $\pi$ electron along with on-site Hubbard 
interaction U.\cite{majumder-U}
\bea
H_{PPP}~=-~ \sum_{ij,\sigma}~t_{ij}~(\hat a_{i~\sigma}
^\dagger \hat a_{j~\sigma}^{} ~+~H.~c. )~+
~\sum_{i}{\hat n_i}{\epsilon_i}~\nonumber\\
~+\frac{1}{2}~\sum_{i}~U_{i}{\hat n}_{i} ({\hat n}_{i} -1)~
+~\sum_{i,j}~V_{i,j}~ ({\hat n}_{i} - z_i)
({\hat n}_{j} - z_j)~
\eea
\noindent
 where $\hat a_{i,\sigma}^\dagger(\hat a_{i,\sigma})$ creates (annihilates) a
 $\pi$ electron in the $p_z$ orbital on atom site $\textit i$ with spin $\sigma$.
 $n_i=\sum_{\sigma}n_{i,\sigma}$ is the total no. of electrons on atom $\textit i$.

 $\epsilon_{i}$ is the site energy and $z_{i}$ is the local chemical potential which is one for carbon atom.
 $t_{ij}$ is the long range hopping matrix element (transfer integral) for bonded pair of atoms $\textit i$ and $\textit j$.
 $U_i$ is the on-site Hubbard repulsion term and $V_{ij}$ is the inter-site
 coulomb repulsion term can be parametrized under Ohno's parametrization scheme.\cite{ohnopara}

\begin{equation}
V_{ij}=14.397\Bigg[\Bigg(\frac{28.794}{U_i+U_j}\Bigg)^2+r_{ij}^2\Bigg]^{-\frac{1}{2}}
\end{equation}
 For carbon atom $U_C$ is 11.26 eV and $t_{ij}$ is -2.4 eV for C-C bond length 1.397 (\AA) in benzene.
For bond length alteration the hopping matrix element $t_{ij}$ can be extrapolated using the relation
\begin{equation}
t_{ij}=-2.4\left(1-\frac{\delta_{ij}}{r_{ij}}\right)
\end{equation}
where $\delta_{ij}=(r_{ij}-1.397) \AA$ and $r_{ij}$ is the C-C bond length in test molecules. The atomic site energy for 
carbon atom $\epsilon_{i}$ is 0 eV.
The PPP Hamiltonian can conserve total spin. Thus spin symmetry can be used 
for reducing the dimensionality of the Hamiltonian matrix and hence the 
computational effort. All the systems described here possesses inherent $C_2$
symmetry along the axis perpendicular to the molecular plane.
For systems possessing $C_2$ symmetry the ground (A) state and first dipole allowed state (B) are in even spin parity subspaces.
The dipole-forbidden, two photon state (2A) has also even spin parity.
The triplet states lie in the odd spin parity subspace.
This spin parity can be used to calculate the low lying excited states for both
singlet and triplet energy sub-spaces. Here we have employed spin adapted
 Slater basis for the first four site of these systems.\cite{slater-basis}
The resultant basis is non-orthogonal but linearly independent. Thus in this basis the Hamiltonian matrix will be non-symmetric. Rettrup's modification and Davidson algorithm is employed to obtain few excited states. All the assumptions are implemented in the Hamiltonian (PPP). \\
Here we employed symmetrized Density Matrix Re-normalization technique (DMRG). Density matrix can be formulated considering symmetries.
In DMRG scheme initially we started with four sites and the Hamiltonian matrix is exactly diagonalizable under PPP Hamiltonian scheme. All those molecules are built up by considering two sites at a time at the interior. The connectivity schemes are shown in Figure(\ref{iSF_connect}). All the other structures can be constructed from these schemes. To calculate the low-lying excitations (optical) it is useful to introduce spatial symmetry as well as spin symmetry. The molecules we studied have inherent $C_2$ symmetry about an axis perpendicular to the plane of the molecules as stated before and we employed spin parity symmetry. Later we split-up the Hilbert space into even and odd spin parity. The ground state is in $\textit {A}$ subspace and in ${^e}A$ spin parity. The optically allowed state which is strongly coupled to the ground state lies in the even parity subspace ${^e}B$ and it is called one photon state. The dipole forbidden two photon state is also in the even parity subspace 2${^e}A$. The triplet states are in odd parity sub-spaces. We first calculated the Density of these (A and B) sub-spaces. Average of these matrices were used to produce highest number of eigenvalues and eigenvectotrs in re-normalization scheme.

 All the structures are optimized in $\textit gausian09$ using Density Functional theory within B3LYP exchange correlation functional and 6-31g(d,p) 
basis set.\cite{gauss,becke,lyp} The optimized geometries are shown in Figure(2). \\
Along with the  DMRG calculation we have carried out Time Dependent Density Functional Theory (TD-DFT) calculation for these molecules. We have consider the first 10 excited states and calculated both $S_0-S_1$,  $S_0-T_1$ gap and oscillator strength (f)  and from there we calculated desired energy gaps to compare the results obtained from  DMRG calculation.
\clearpage
\section{RESULTS AND DISCUSSION}
We calculated the low-lying excited energy ordering of indeno fluorene . To observe iSF we form the dimer of these molecules
 through benzene bridge and compared the results obtained in DMRG scheme for the singlet dimers (D) to the triplet state 
energies of the monomers (M). Next we have calculated the excited energy of these systems (both monomer and dimer) in time 
dependent density functional (TDDFT) method.\\
The calculations showed that due to the dimerization the absorption maximas for the systems fall subsequently $\sim$0.5 eV.
The first excited triplet energy is 0.84 eV. Thus the thermalization loss can be minimized. In case of xSF the crystal
packing plays a major role as the electron transfer rate constant depends on the molecule to molecule distance in a
molecular crystal. The covalent benzene spacers did not change energy of the first excited state of the dimer with no benzene
spacers much. \\
The frontier molecular orbital calculation shows for triplet the HOMO and LUMO consisting of localized orbitals. Which 
indicates the coupled $^1(TT)$ state. In case of xSF this state is localized of two adjucent chromophore. But in this case
this state is localized on a single molecular dimer in either side of the covalent bridge. This $^1(TT)$ state is the 
seed for efficient singlet fission and formation of this state discard the possibility of triplet-triplet annihilation.
As this state is non-fuluorescent it is very difficult to find out this state. Recently there are some spectroscopic
signatures of the state are found experimentally while observing transient absorption spectrum for different molecules\cite{yong2017entangled}.
In a review in 2018 by Kim and Zimmerman discussed the necessity of consideration of this coupled state in detail.\cite{kim2018coupled}
\begin{table}[h]
\caption{\bf Computed values for indeno-fulorene dimers}
\begin{tabular}{ccccccc}
\hline
\hline
Molecule & Method & $S_0-S_1$ & $2A_g$ & $S_0-T_n$ & $\mu$ (debye)/f \\
 &  &(D)(eV) &(D)(eV) &(M)(eV) & \\
\hline
  & DMRG & & & & &  \\
  &      &2.14 &1.90 &0.84 &0.82 \\
  &      &3.02 &     &1.84 &0.09 \\
  &      &3.07 &     &     &4.27 \\
  &      &3.28 &     &     &0.58 \\
  &      &3.45 &     &     &2.98\\ 
$\textbf{B0}$ &      &  & & & \\
  & TD-DFT & &  & & \\
  &      &1.89 &1.91 &0.90 &0.03 \\
  &      &2.24 &     &2.10 &0.00 \\
  &      &2.24 &     &     &0.00 \\
  &      &2.43 &     &     &1.12 \\
  &      &2.49 &     &     &0.01 \\
\hline
  & DMRG & & & &  \\
  &      &2.20 &1.83 &0.86 &0.05 \\
  &      &2.91 &     &1.84 &0.05 \\
  &      &3.43 &     &     &0.05 \\
  &      &3.62 &     &     &0.09 \\
$\textbf{B1}$ &       & & & & \\
  & TD-DFT & & &  & &  \\
  &      &1.91 &1.92 &0.90 &0.03 \\
  &      &2.32 &     &2.10 &0.01 \\
  &      &2.32 &     &     &0.00 \\
  &      &2.45 &     &     &1.20 \\
\hline
  & DMRG & & & &  \\
  &      &2.28 &1.85 &0.86 &0.19 \\
  &      &3.26 &     &1.84 &0.05 \\
  &      &3.75 &     &     &0.05 \\
  &      &3.86 &     &     &0.09 \\
  &      &3.90 &     &     &0.05 \\
$\textbf{B2}$ &       & & & & \\
  & TD-DFT & & & & \\
  &      &1.91 &1.92 &0.90 &0.03 \\
  &      &2.36 &     &2.10 &0.00 \\
  &      &2.36 &     &     &0.00 \\
  &      &2.45 &     &     &0.07 \\
  &      &2.46 &     &     &0.54 \\
\hline
\label{table1}
\end{tabular}
\end{table}
\clearpage
\subsection{Comparison with experimental results}
The computed values in both DMRG and TDDFT scheme are in exelent agreement with the experimental results. In the year
2010 Jinseck Kim $et\ al.$ sythesized conjugated copolymers based on indeno-fluorene skeleton. 
They synthesized four co-polymers named PIF-DBT35, PIF-DBT50, PIF-DTP35 and PIF-DTP50, where IF reffers indeno-fluorene, and found their absorption maxima in chlorofom
solvent at 547 nm (2.26 eV), 550 nm (2.25 eV), 638 nm (1.94 eV) and 645 nm (1.92 eV) respectively\cite{indeno-expt1}. These 
values are in excellent agreement with the computed DMRG and TDDFT values for $S_0-S_1$ as given in table:\ref{table1}. 
Recently Ying Sun $et\ al.$ also sythesized indeno-fluorene based conjugated copolymers. The three co-polymers named PIDFDQ, 
PIDFDPP and PIDFPyT, where IDF reffers indeno-fluorene, and found their absorption maxima in chlorofom
solvent at 551 nm (2.25 eV), 668 nm (1.85 eV) and 561 nm (2.21 eV) respectively\cite{indeno-expt2}. These values are in 
excellent agreement with the computed DMRG and TDDFT values as given in table:\ref{table1}.
 The experimental observation also shows the absorption maxima does not change much in thin film rather solution\cite{indeno-expt1,indeno-expt2}.\\
Unfortunately we did not found any triplet excitation energy for indeno fluorene monomer.
\clearpage
\section{CONCLUSION}
We conclude that indeno[1,2-b]fluorene which poseses xSF, can also applicable as iSF materials. The localization of triplet
frontier orbitals are signature of coupled $^1{TT}$ state which can dissociates to the inorganic solar cell materials. The
sudden surge in the charge carrier density increases the current. Hence the SQ limit, which is a theoretical barrier for
efficiency of single junction solar cell can be underestimated. The advantage of iSF is that there is nothing to worry
about the crytal packing which plays a important role in charge transfer in xSF process. Our calculation shows that
the observed absorption maximas are very close to the experimental observation. Thus indeno[1,2-b]fluorene can be utilized
both as xSF and iSF materials.
\section{ACKNOWLEDGMENT}
Authors acknowledge financial support from Department of Science and Technology,
Government of India through a SERB Fast-Track Grant
SB/FT/CS-164/2013 and IISER Kolkata.
\bibliographystyle{unsrt}  
\bibliography{iSF_indeno-fluorene}

\newpage

\end{document}